# *Sub-Electron Read Noise at MHz Pixel Rates

Craig D. Mackay, Robert N. Tubbs, Institute of Astronomy, University of Cambridge,
Madingley Road, Cambridge, CB3 0HA, UK

Ray Bell, David Burt, Ian Moody
Marconi Applied Technologies, Chelmsford, Essex, UK

## ABSTRACT

A radically new CCD development by Marconi Applied Technologies has enabled substantial internal gain within the CCD before the signal reaches the output amplifier. With reasonably high gain, sub-electron readout noise levels are achieved even at MHz pixel rates. This paper reports a detailed assessment of these devices, including novel methods of measuring their properties when operated at peak mean signal levels well below one electron per pixel. The devices are shown to be photon shot noise limited at essentially all light levels below saturation. Even at the lowest signal levels the charge transfer efficiency is good. The conclusion is that these new devices have radically changed the balance in the perpetual trade-off between readout noise and the speed of readout. They will force a re-evaluation of camera technologies and imaging strategies to enable the maximum benefit to be gained from these high-speed, essentially noiseless readout devices. This new LLLCCD technology, in conjunction with thinning (backside illumination) should provide detectors which will be very close indeed to being theoretically perfect.

## 1. INTRODUCTION

Although CCDs are suitable for a very wide range of applications there are still a number of areas where CCDs show significant limitations in their performance. The principal weakness is that their readout noise (the system noise that is achieved in the absence of any input signal) increases substantially as the pixel readout rate increases. As many applications are demanding increasing resolution, it is essential that the corresponding increase in pixel rate is not accompanied by a reduction in performance caused by increasing readout noise, particularly when the full well capacity of the higher resolution CCDs (generally with smaller pixels) is also significantly reduced.

A novel CCD architecture has been developed by Marconi Applied Technologies, Chelmsford, UK with a view to providing CCD cameras with a sensitivity and readout noise similar to those obtained by the best image intensifier. This same architecture will also allow a dramatic improvement in the performance of high-speed scientific imaging systems for a variety of applications.

An evaluation of a video rate camera using a low light level CCD (LLLCCD) has already been published by Harris et al. and a more detailed description of the architecture of the device is also presented by Cochrane et al. (2000), by Wilson (2000) and at this conference by Jerram et al. The purpose of this paper is to carry out the critical evaluation of the LLLCCD technology developed by Marconi Applied Technologies so as to quantify in is much detail as is possible the performance of the LLLCCD technology for scientific imaging applications.

## 2. LLLCCD ARCHITECTURE

The technology behind the LLLCCD is disclosed in the European patent application EP 0 866 501 A1. In essence a conventional CCD structure is used with the output register extended with an additional section that has one of the three phases clocked with a much higher voltage then is needed purely for charge transfer. The large electric fields that are established in the semiconductor material beneath pairs of serial transfer electrodes cause charge carriers to be accelerated to sufficiently high velocities that additional carriers are generated by impact ionisation on transfer between the regions which are under the electrodes. The charge multiplication per transfer is really quite small, typically one percent but with a large number of transfers (591 for the device characterised here) substantial electronic gains may be achieved. The output of this extended serial register is passed on to a conventional CCD output amplifier. The electronic noise of this amplifier which might be equivalent to a few tens of electrons at MHz pixel rates is now divided by the gain factor of the multiplication register which, if this gain is high enough, will reduce the effective output read noise to levels much smaller than one electron rms.

*Correspondence: email: cdm@ast.cam.ac.uk

This architecture has many advantages. All the developments that had led to the astonishingly high performance of scientific CCDs such as their remarkable charge transfer efficiency, the extremely high quantum efficiencies of thinned (back illuminated) devices and the very low dark currents that are now achieved by operating the imaging area in inverted mode are unaffected by the high gain multiplication register of the LLLCCD. It will also be clear that by varying the amplitude of the higher voltage clock phase in the extended register the net gain of the register may be varied from unity (when the multiplication register is operated with normal clock levels, and the output amplifier will give its normal equivalent readout noise) to a high gain which could be in excess of 10,000. The only limitation of this method is that the dynamic range of the CCD operated in high gain will be limited by the capacity of the multiplication register in electrons divided by the gain of register.

## 3. SIGNAL-TO-NOISE CONSIDERATIONS

It is relatively straightforward to model the characteristics of the multiplication register. This register has 591 stages (for the CCD 65 discussed here) each of which offers a low probability ( p, typically 1%) of converting one electron into two electrons. The overall gain is $(1+p)^{591}$. In this way a multiplication probability of 1% will give an overall gain of approximately 358, while a gain probability of 1.5% will give the mean gain of 6670. In an ideal world the same gain value would be applied to every electron which enters the multiplication register. Unfortunately because of the statistical nature of the multiplication process there is a wide dispersion in the number of electrons generated, from one input electron. This is shown in figure 1 where for a gain probability of 1.5% (overall gain of 6670) is shown the distribution of the number of output electrons per input electron. The probability distribution shown in [Figure 1] was obtained by convolving together appropriate probability distributions for the amplification stages.

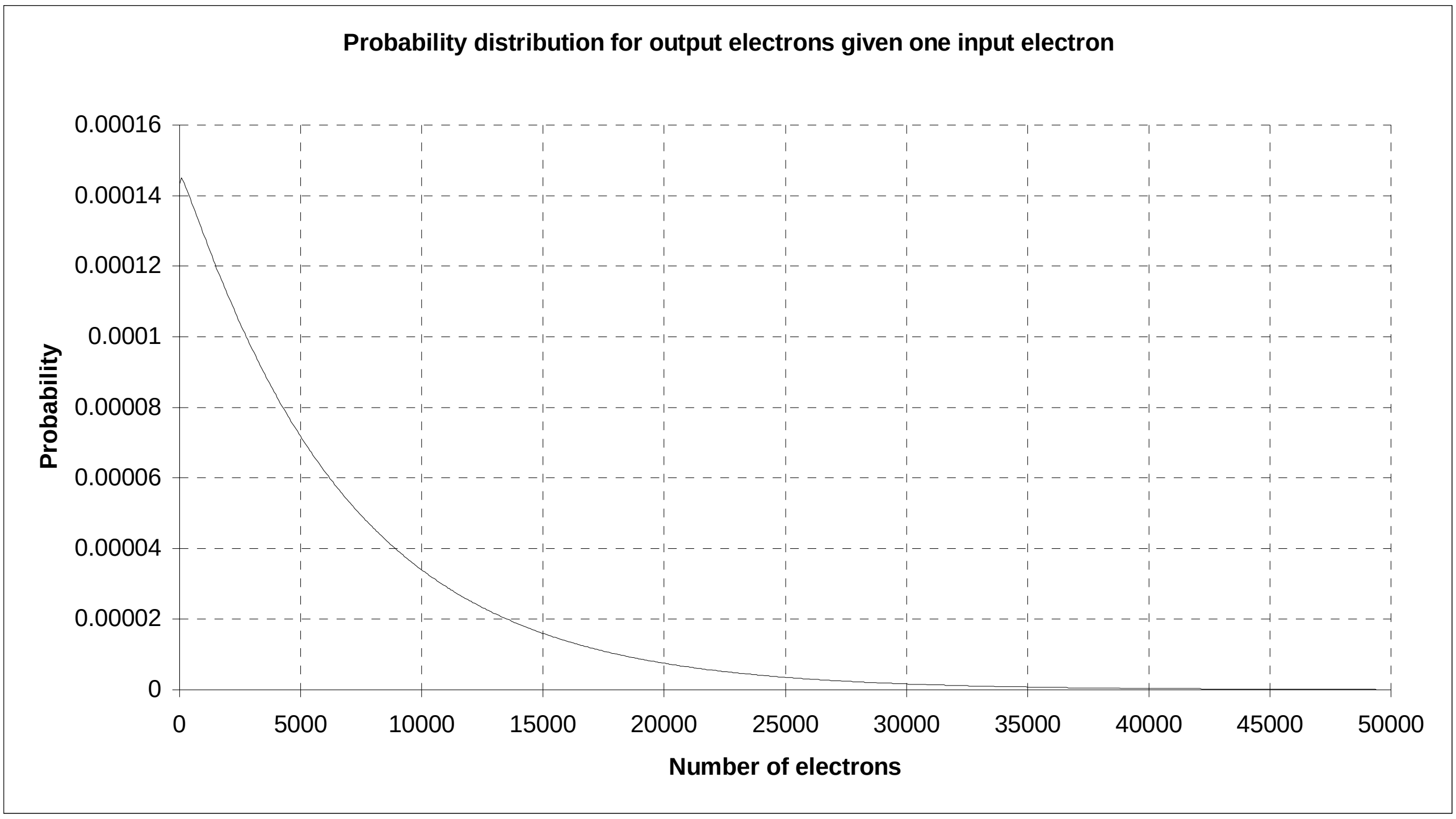

Figure 1: The probability distribution for the signal output from a multiplication register of 591 stages with a mean gain per stage of 1.015. The mean overall gain for the entire multiplication register is then 6670.

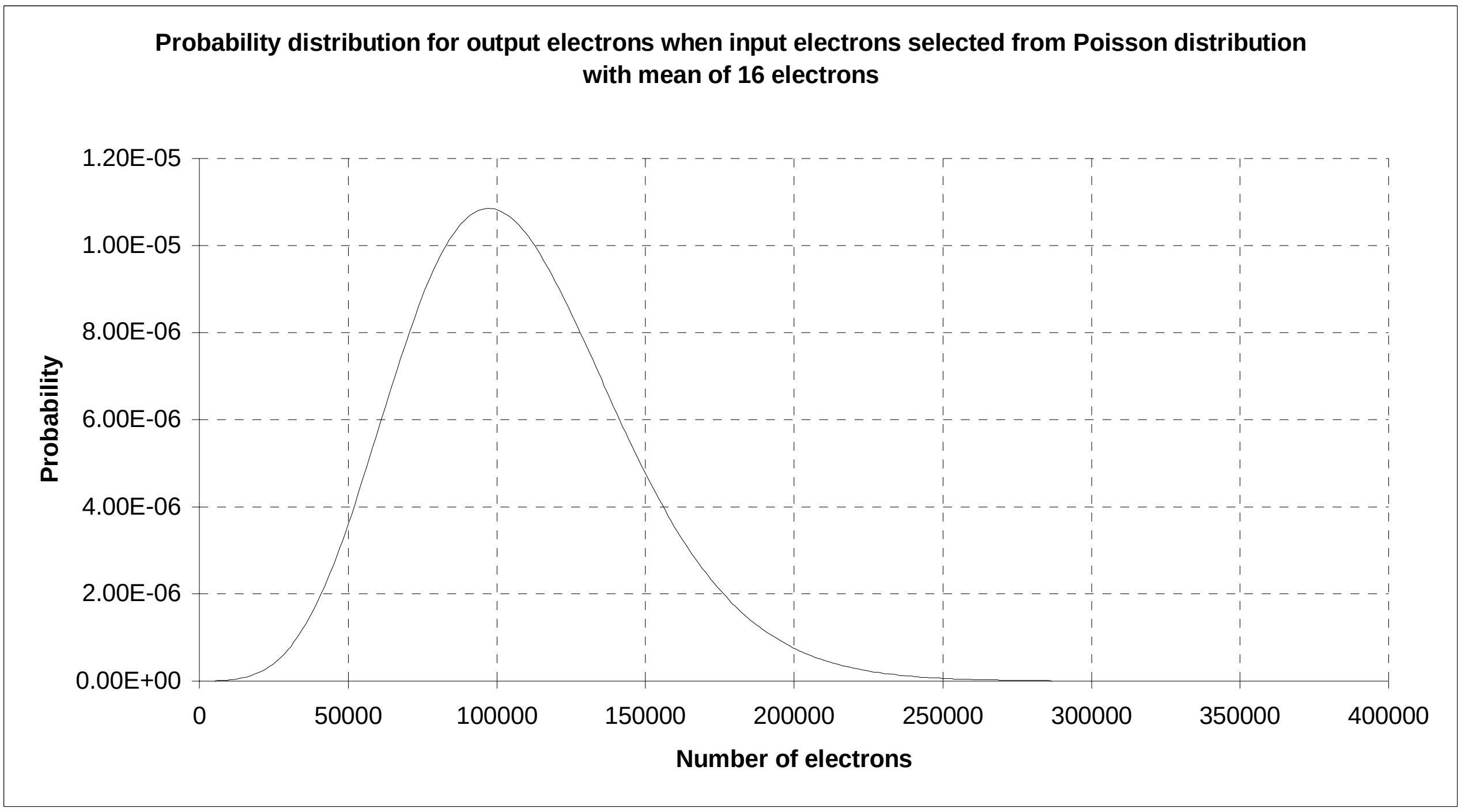

Figure 2: The probability distribution in the number of electrons output from a multiplication register of 591 stages with a mean gain per stage of 1.015, and a total overall gain of 6670 with an input signal which has a mean of 16 electrons and a Poisson distribution.

We can see that this distribution is a monotonically decreasing function with no peak value around the average gain value. The standard deviation in the gain that we get is virtually identical to the mean. The effect of the gain can be seen in Figure 2 which shows the number of electrons coming out of the multiplication register. When there is no gain in the register we have a mean number of electrons equal to 16 and an RMS dispersion of 4 electrons. The distribution in the number of electrons output from the multiplication register when operated with a high gain (>10) has a mean value of 16 times the gain, but the signal-to-noise in the output electrons is reduced by a factor of $\sqrt{2}$. This reduction in signal-to-noise is brought about by the dispersion in the gain of the multiplication stages. We can therefore identify two regimes: firstly if the gain is unity, then the signal-to-noise in the output of the register is equal to the shot noise, whereas if the gain is high then the signal-to-noise is decreased by a factor of $\sqrt{2}$. However, when the gain is high the effect of the readout noise will be insignificant, providing an enormous improvement at low light levels. This reduction in the signal-to-noise for a given number of input electrons is exactly equivalent to the effect of using a detector system free from this effect but which has a detective quantum efficiency exactly half that of the device operated in unity gain mode.

We can also see that it is possible to run the device in a photon counting mode as follows. Standard CCDs cannot be used for photon counting since even at slow read out rates they have a read noise of 2-3 electrons minimum. With the LLLCCD technology we can see that if we set the multiplication register gain to be much higher than the device readout noise in the absence of gain then the great majority of electrons entering the multiplication register will exit with an amplitude which is very much greater than the noise in the output stage. Selecting threshold of, say, five times the standard deviation in the noise of the output amplifier allows essentially all the electrons entering the register to be detected with a good signal-to-noise. Because of the wide dispersion in the energies of these amplified electrons the best way to process them is to accept that each represents one photon and to ignore the dispersion. We give each event detection a weight of unity. In this way we are able to restore all the lost quantum efficiency by working in photon counting mode. The big disadvantage is that we have to restrict the signal intensity so that there is an acceptably small risk of two photons been detected on the same pixel. Should this happened then two photons will be counted as one, something we cannot discriminate against because of the monotonically decreasing pulse high distribution shown in figure 1. In practical terms it means that the maximum photon rate that can be tolerated without significant non-linearity is approximately one photon per pixel in 30 frames. At this rate then 30 photons will be counted as 29 photons. By looking carefully at the Poisson statistics of a photon stream we can show that 1 photon in 50/30/10 frames gives 1/1.6/5% non-linearity. This non-linearity is very predictable and can be corrected for, given the locally detected photon rate. It does, however, give a small reduction in detective quantum efficiency which at higher photon arrival rates will become significant and lower the resulting DQE to the levels that are seen with the gain mechanism working normally as described above. This coincidence loss is exactly the same for other intensifier based photon counting systems that use framing read-out systems such as the electron-bombardment CCD, or systems that use an intensifier with video camera read-out with photon counting hardware. The main difference here is that the CCD is a relatively low-cost, solid-state device with higher quantum efficiency when back-illuminated with essentially infinite life and immunity from light overload. In fact the coincidence losses with phosphor based image intensifiers can be poorer because the intensified photon event can cover several read-out system pixels, greatly increasing the likelihood of coincidence occurring. The frame rate limitation may not be a problem, depending on system design. A CCD with multiple outputs can allow very fast frame rates as can systems that read out a smaller sub-array of the CCD. For smaller areas the frame rates can be much higher, with the CCD65, for example, reaching 500Hz frame rate for a 128x128 sub-array.

It is also important to appreciate the importance of deep cooling if photon counting work is to be attempted. The CCD65 works in inverted mode to give the very low dark current level of 200 electrons/px/sec at +20C. This is virtually eliminated by cooling to -140C. Cooling to a typical Peltier-cooled temperature of -40C will reduce the dark current to about 1 electrons/pixel/second, far too high for photon counting rates where the maximum for linear operation might be 0.01 to 0.1 electrons/pixel/second from the signal plus the dark rates together.

So we now have three different regimes in which it is possible to operate this device:

1. The conventional CCD mode, with no gain in the multiplication register, and the signal-to-noise set by the photon shot noise added in quadrature with the readout noise of the CCD output amplifier.

2. The CCD operated with a gain in the multiplication register that substantially overcomes the readout noise of the output amplifier. In this case the signal-to-noise is worse than would be expected from the number of photons detected by a factor of root 2. Another way to think of this degradation is to calculate on the assumption that the signal-to-noise is set by the photon shot noise but that the detector has half the detective quantum efficiency that it has been for mode 1 above.

3. The CCD is operated with high gain in the multiplication register so that the readout noise of the CCD output amplifier is completely negligible for each multiplied electron. If each event is then thresholded and treated as a single event on equal weight without making any attempt to consider its amplitude then the quantum efficiency that is lost by operating in mode 2 above is restored, giving the same quantum efficiency essentially as that in mode 1 above. There are, however, the major limitations that the maximum photon rate used must be kept extremely low in order to avoid coincidence losses which will give rise to non-linearities in the response curve of the detector system, and the corresponding need for deep cooling to maintain correspondingly low levels of dark current.

It is worth noting that modes 2 and 3 above are not mutually exclusive. By designing a system with parallel output signal electronics, that can photon count as well as frame average it is possible to imagine systems that are photon counted in those parts of the image where the signals are low enough to give good linearity, and where the ultimate DQE is important, while in those parts of the image that are brighter, the negligible read-out noise mode of operation is also achieved, and frame averaging gives the full dynamic range.

## 4. THE MARCONI CCD65 ARCHITECTURE

The tests described in the remainder of this paper were carried out with a CCD 65 manufactured by Marconi Applied Technologies, Chelmsford, UK. The CCD 65 is designed for frame transfer interlaced operation for use in cameras operating at PAL or NTSC video rates. The image area consists of 576(H) by 288(V) pixels each of 20 by 30 microns, and works in inverted mode to give an extremely low dark current (typically 200 electrons per pixel per second at a temperature of 20 C. The image and store areas use a 2-phase structure with interlace provided in video operation by deriving alternate frames after integration in inverted mode (with both phases low) by bringing up one phase or the other. The parallel registers allow transfer rates of up to about 1 MHz and are non-antiblooming giving a 200,000 electron full well. The readout register is able to operate at pixel rates up to in excess of 12 MHz. The multiplication register has a full well capacity of approximately 900,000 electrons in order to permit high gain to be used without saturation.

The output amplifier has a responsivity of 1.3 microvolts/electron and is very similar in design to the large signal, high-speed scientific type output amplifier which is included on devices such as the Marconi CCD 55. The imaging area has a full well capacity of 200,000 electrons. Although this is not a format of CCD that would normally be chosen for scientific imaging applications it is built using this same fundamental architecture that is common to many other devices in the Marconi range. The principal difference from their standard devices is, of course, in the multiplication register. The approach, therefore, in this paper was to characterise the device in its normal mode of operation (no additional gain from the multiplication register) and then see to what extent those characteristics were modified as a function of gain in the multiplication register.

## 5. INSTRUMENTAL CONFIGURATION

The tests were done with a conventional Capella 4100 CCD imaging system manufactured by AstroCam Ltd (now PerkinElmer Life Sciences Ltd). The system is fully programmable and may be run at pixel rates of up to 5 MHz with 14 bit digitisation. The system has programmable gain, clock waveform generation and signal processing timing. It is fully integrated with several software packages that allow an accurate quantitation of many of the properties of CCDs. The CCD65 loaned by Marconi was mounted both in a compact thermoelectric head which allowed the CCD to be operated at approximately -30 C and also in a liquid nitrogen cooled dewar that operated at approximately -140 C so that all sources of dark signal could be eliminated. This was found to be necessary because there was some evidence that there was excess dark current generated in the multiplication register, possibly because of the high electric fields in it.

An additional circuit board was provided to generate the higher voltage clocks needed for the multiplication register. Because the system is fully programmable it was necessary to design a driver that followed the high-speed clock generated by the Capella 4100 controller as closely as possible and gave an adjustable clock high-level that varied between 6 V above substrate (the level used in the standard output register and therefore the level that provided no gain in the multiplication register) and the maximum level in excess of 40 volt above substrate, in order to provide the maximum possible gain from the CCD multiplication register. With careful design it was possible to produce drivers capable of 50 volt slew in approximately 20 ns. One important design issue is that the gain (see later) is critically dependent on the clock voltage. If the stability is to be achieved when the gain is high then it is essential that the clock levels used are stable to only a few millivolts. In all other respects the electronics used were completely standard. The measurements were all carried out at a pixel rate of 1 MHz.

The performance of the CCD 65 system was checked with a wide range of light wavelength, from a calibrated projector that used a variety of LEDs working between 470 and 950 nm.

## 6. TEST RESULTS

The CCD 65 was fully characterised at room temperature, Peltier cooled to -30 C and liquid nitrogen cooled to -140 C with the multiplication register gain set to unity. In all respects the device behaved like a completely standard, normal Marconi CCD. The CCD 65 used clock voltages and waveforms which were similar to those used by current Marconi CCD families such as the CCD 55.

Careful measurements were made of the gain achieved from the multiplication register as a function of the high voltage level used on the phase in the multiplication register that provides the gain. These results are shown in figure 3 which shows both the gain and the readout noise of the CCD as a function of gain clock voltage. Gains were measured as high as 45,000, and the corresponding readout noise was measured to be as low as 0.002 electrons rms. The fact that such measurements were made is not intended to imply that it would be in any way practical to operate CCD 65 at this gain level but rather that it is possible to measure gains as high as this.

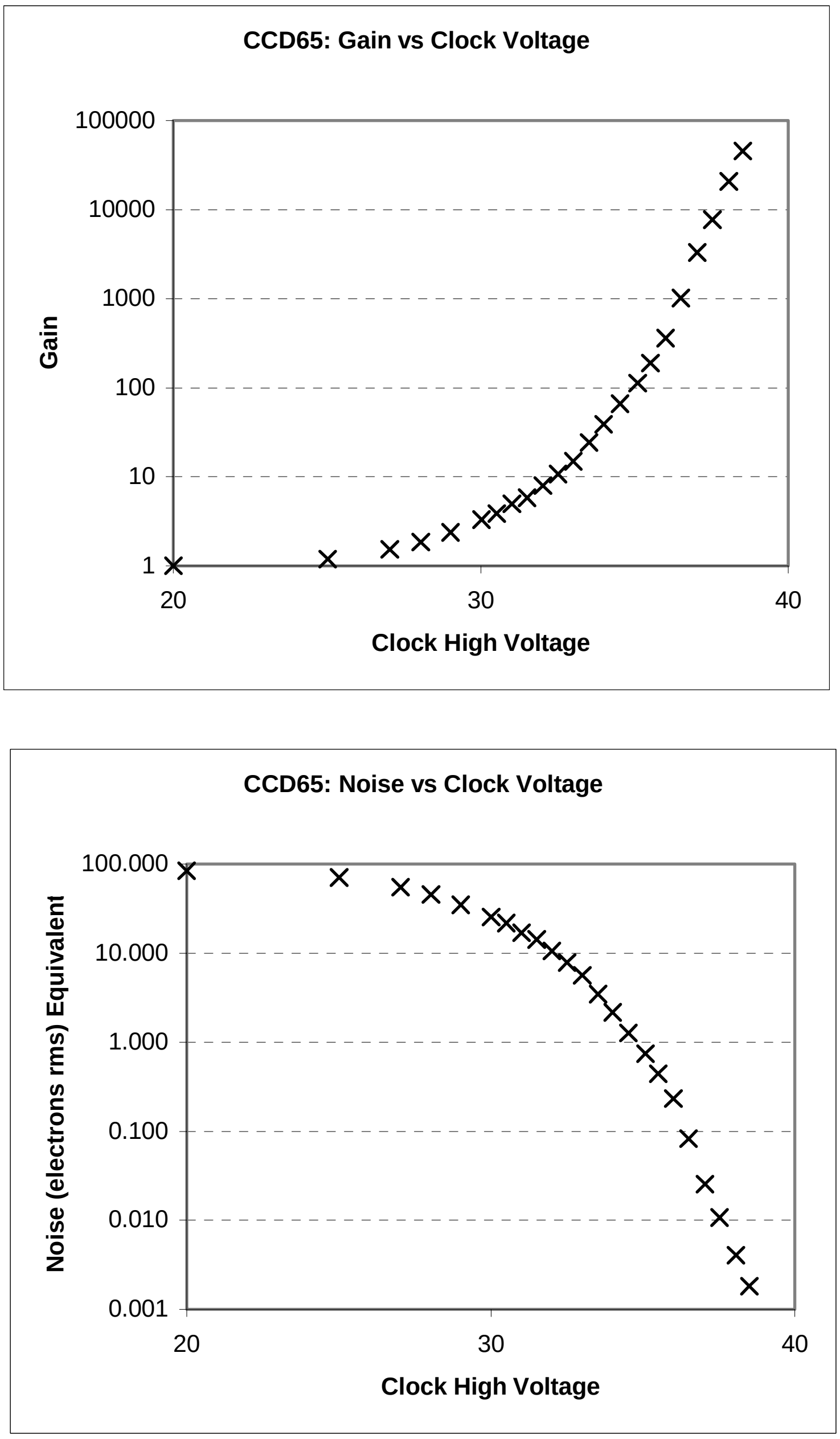

Figure 3: The effect of changing the clock high level (relative to the substrate voltage) in the multiplication register on the gain and hence on the effective read-out noise of the CCD65 device tested here. The gain increases very rapidly for a relatively small change in clock high voltage, showing that gain stability will only be achieved by excellent clock-high voltage stability. At gain levels around 1000, the gain triples for 0.5 volts clock high increase, making it necessary to achieve millivolt clock-high repeatability for 1% gain stability. As the achieved gain comes from the plots shown convolved with the clock waveforms, careful control of clock waveform ringing repeatability is also essential. These results were obtained at a CCD temperature of approximately -30C.

What is clear from the results shown in figure 3 is that the gain increases very rapidly indeed with slight increases in the clock voltage. The consequence of this is that it is essential to make clock drivers that are extremely stable and have very low levels of noise on them. The clock drivers designed for the system described here had a stability of better than one millivolt.

It was also noticed that the gain that derived from a specific voltage level varied significantly with temperature. Between +30C and +12C, the gain approximately doubles for a 9C drop in temperature. The difference measured can be expressed by saying that the gain achieved at a specific voltage at -30 C was achieved with a voltage lower by 2.3 V at -140 C.

One of the most demanding tests to make on the system is to compare a single image taken at a specific light level with another image obtained by adding a large number of individual images each taken at a much lower light level so that the total number of photons per pixel in the two images should be the same. This was done at a number of light levels and the results are entirely consistent with the descriptions given above. Examples of these images are shown in figure 4. These images were taken at a gain level of about 3350 using blue (470nm) light. The images show that the CCD is able to transfer charge perfectly happily at these low signal levels, and that the image quality depends only on the integrated signal level and not on the number of images added together to achieve this signal level. The only difference is that there is a greater level of white spots on the image. They can be seen most clearly in the middle image of the left hand sequence of Figure 4. It is not clear what their origin might be.

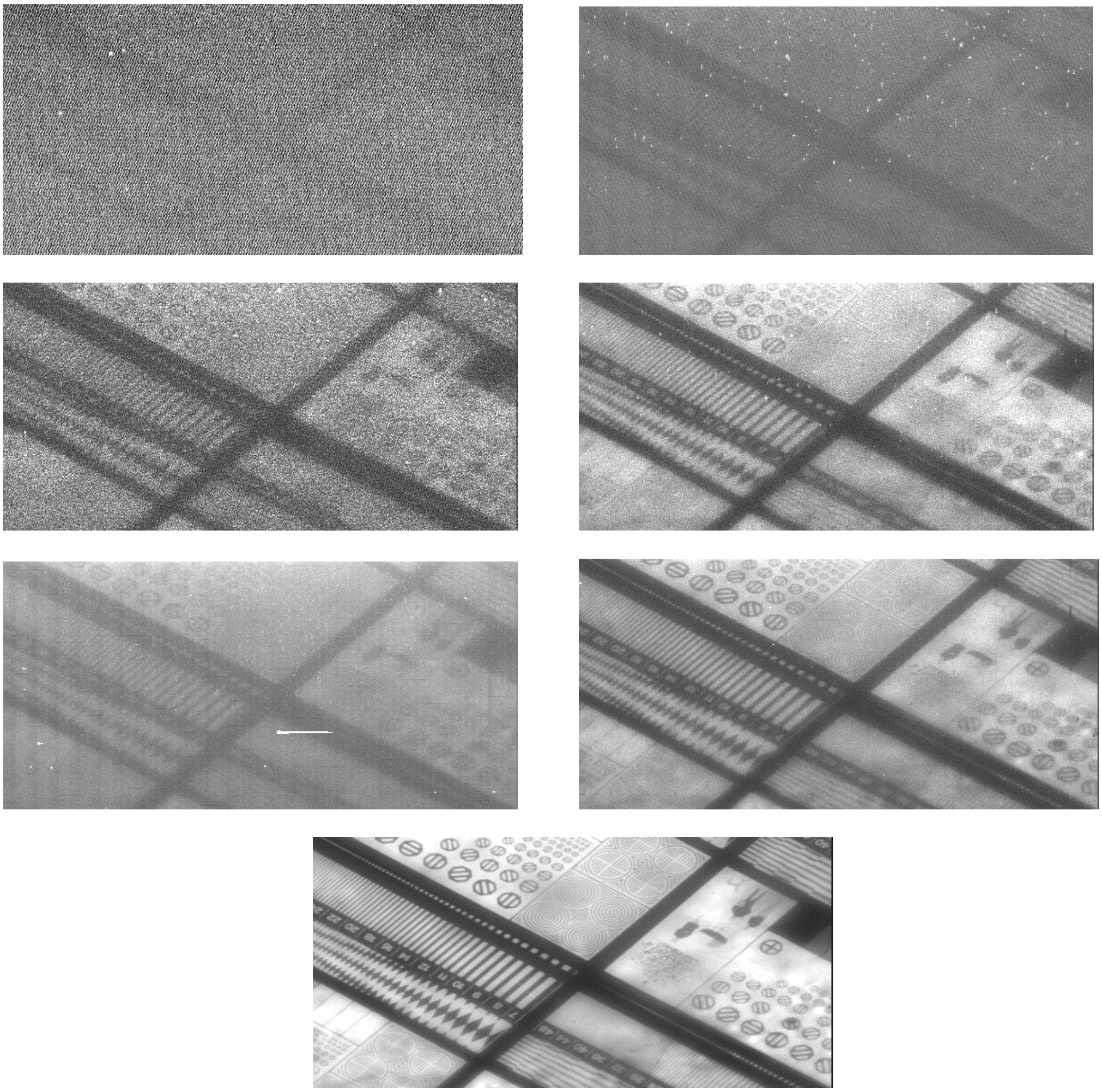

Figure 4: A series of test chart images showing that the CCD65 does allow effective operation at extremely low signal levels. The sets of three images on the left and the right show a single exposure (top), the sum of 16 exposures (middle) and a sum of 256 exposures (bottom of the three) at peak signal levels of approximately 1.5 detected photons per pixel per frame (left series) and 20 electrons per pixel per frame (right series). Underneath these is an image taken at much higher signal levels to show the appearance of the images being used.

## 7. REASSESSING APPLICATION TRADE-OFFS

There are two main regimes experienced when operating CCD cameras for scientific applications. In the high light level case the photon shot noise is significantly greater than the system readout noise. Here there is no advantage at all in using the LLLCCD technology. In the other low light level regime, the photon shot noise in the image is comparable to or less than the readout noise. The effect is to make the overall system noise significantly poorer than would be expected purely from photon shot noise statistics, something that effectively reduces the overall system detective quantum efficiency. It is in this regime that the LLLCCD technology has a great deal to offer.

In its simplest application, the LLLCCD technology simply allows the readout noise of the system to be reduced to a level where the photon shot noise will always dominate. In order to preserve dynamic range as much as possible it is sensible to minimise the multiplication register gain since that gain value is the same factor by which the dynamic range of the device is reduced. Using LLLCCD technology it is always possible to make a CCD system that is essentially free from readout noise but this is generally at the expense of loss of dynamic range. Of course it is always possible to run at a high frame rate and add together many images in order to extend the overall dynamic range. The potential of LLLCCD technology in providing much higher gain than is usually possible allows much faster readout rates to be used without losing the essentially noise free capability of this technology. The faster frame rates will inevitably reduce the dynamic range per image. However images can be added after readout to give the dynamic range necessary and this may be acceptable in some applications.

However it is very important when considering the use of LLLCCD technology in any application to realise that many of the assumptions about CCDs (and in particular that it is essential to run them as slowly as possible in order to minimise the readout noise) are inappropriate if you have enough gain to overcome the intrinsic noise of the amplifier and produce an overall system readout noise that is negligible.

A good example is the use of CCD detectors as wavefront sensors for applications in adaptive optics. In these applications it is necessary to measure the phase errors that affect the flatness of the light wavefront that comes into a telescope. The phase errors are created by turbulence in the atmosphere and are responsible for the loss of spatial resolution in images detected from the ground. A common strategy for measuring these wavefront errors is to use a Shack Hartmann arrangement whereby sub-apertures of the telescope are imaged separately using an array of lenslets. In the image plane of the Shack Hartmann sensor there is an array of stellar images formed, each image from one lenslet which covers one of the sub-apertures of the telescope (figure 5). Each star image is tracked as it moves around in response to the phase errors across the sub aperture. The amount of image motion is converted back into a phase error pattern across the whole telescope aperture and a flexible mirror is distorted to compensate for these errors so as to give an error free and therefore diffraction limited image in the telescope image plane.

In order to get to as faint a limiting magnitude as possible the Shack Hartmann system uses the minimum number of lenslets across the aperture as this approach makes each star image as bright as possible. In addition the detector is read out as slowly as possible in order to minimise the readout noise. As a consequence the whole spatial and temporal construction of the system has effectively made assumptions about the scales of the errors to be encountered which generally will be incorrect for the actual night in question.

The use of gain CCD system based on LLLCCD technology allows the detector to be run much faster than is likely to be necessary so that computer software can average the images temporally in a way that can be changed dynamically depending on the correlation times of the atmosphere on the night in question. In this way the effective readout rate may be made faster or slower in response to the real conditions experienced. When conditions are good, the slower readout possible allows much fainter objects to be used. This is important because there are many regions of sky with this technique cannot be used as there are not adequately bright guide stars within the field a view. The fainter the guide star, the larger at the number of objects may be observed.

It is further possible to avoid the restrictions which are placed on the spatial scales of turbulence by the use of a fixed lenslet array. We may use a continuous wavefront sensor such as a shearing or a curvature sensor. By using a higher resolution detector than is strictly necessary to detect the phase errors it is possible to spatially combine the signals in order to give the best description of the phase errors across the telescope aperture. If the conditions are particularly good then an approach like this will allow of the spatial and temporal scales of the detector system to be adjusted in real time to suit the conditions.

Under the best conditions the phase errors change on relatively large scales and they vary relatively slowly. This allows operation at much fainter levels than is possible under the same conditions with a fixed rate and fixed resolution wavefront sensor as is often used, provided the system may be dynamically reconfigured in this way.

Another related application is in "Lucky Astronomy" where large numbers of short exposures are taken at high speed (to beat the natural fluctuations in atmospheric seeing) and sorted to select those with the best images (Baldwin et al, 2001, ref. 3). Normally this would be very inefficient because of the high read-out noise of a fast read-out CCD system, and the low signal levels per frame. The LLLCCD technology completely changes this trade-off, allowing spatial and temporal averaging after read-out, and giving dramatic improvements in limiting sensitivity.

Other applications that will benefit from the use of LLLCCD technology include:

1. Bio- and chemi-luminescence imaging where extremely low light levels often require substantial binning factors to ensure that signal levels are large enough to overcome CCD read-out noise. LLLCCD technology would allow images to be read out unbinned, and selectively averaged depending on the signal levels actually achieved (Mackay, 1999, ref 4))
2. High-speed confocal microscopy (Mackay, 1998, ref. 5) where the need to achieve good signal-to noise and short frame times when working with dynamic systems is often limited by CCD read-out noise and resolution compromises.
3. Astronomical spectroscopy, where the need to take multiple images to give good cosmic ray suppression worsens overall read-out noise. In addition, spectra often are best taken at high resolution to give optimum night-sky suppression and at low resolution to give good signal to noise ration on the faintest parts of the spectrum. The LLLCCD technology again allows selective post-read-out binning to be used.
4. X-ray and neutron beam imaging can also suffer from low signal levels, and from the need to have multiple read-outs to allow discrimination against directly detected X-ray events. The LLLCCD technology will allow intelligent frame averaging to be carried out to allow reliable event suppression as well as minimising read-out noise.

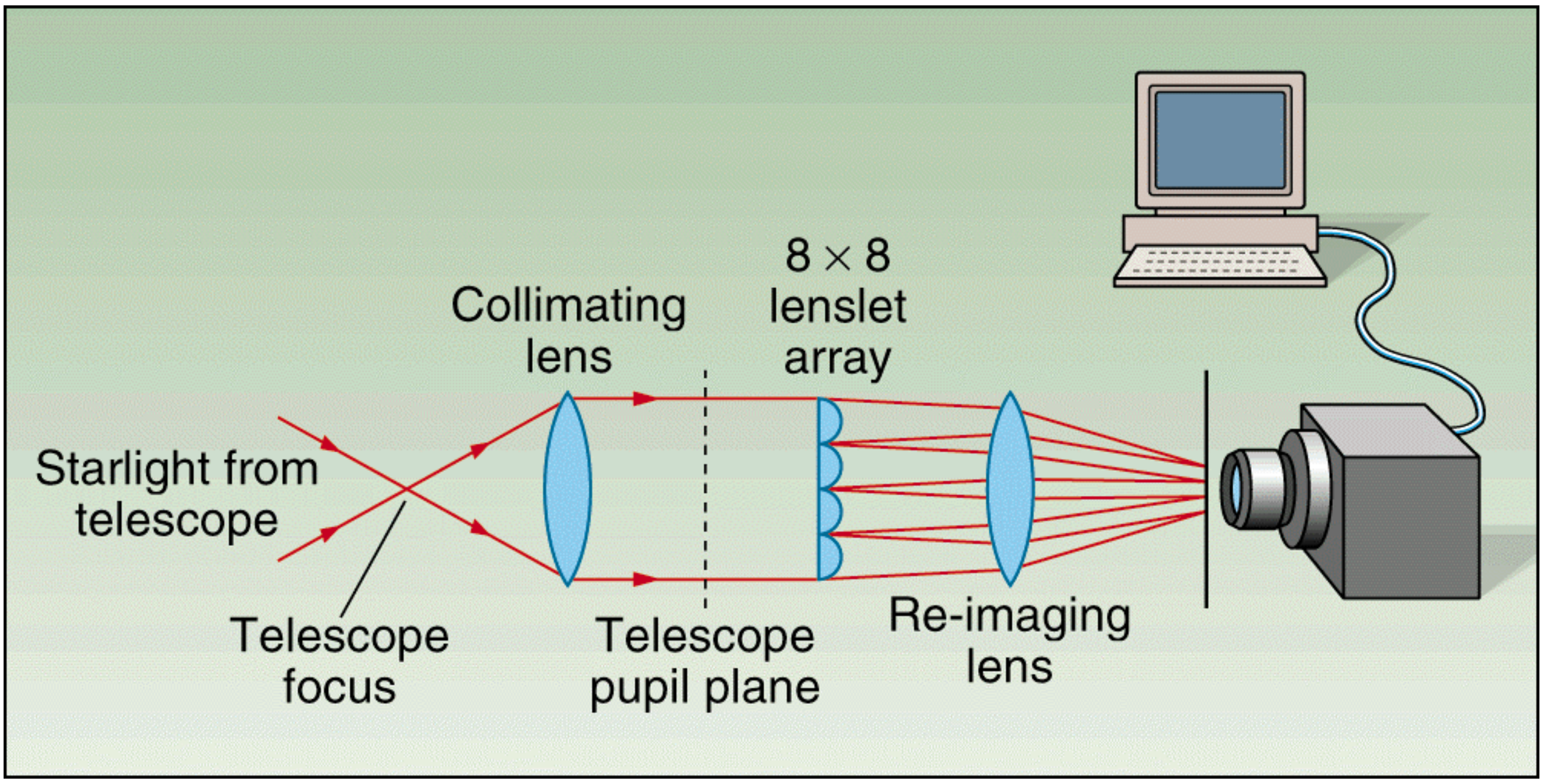

Figure 5: The Shack-Hartmann sensor system layout. The light from a star is reimaged with an 8 by 8 array of sub-apertures onto a CCD camera so that the motions of each can be tracked. (picture courtesy of Laser Focus World)

## 8. CONCLUSIONS

We have undertaken a thorough examination of the performance of a scientific imaging system based on the new LLLCCD technology described above. The three modes of operation have been identified:

1. The conventional CCD mode, with no gain in the multiplication register, and the signal-to-noise set by the photon shot noise added in quadrature with the readout noise of the CCD output amplifier.

2. The CCD operated with a gain in the multiplication register that substantially overcomes the readout noise of the output amplifier. In this case the signal-to-noise is worse than would be expected from the number of photons detected by a factor of root 2. Another way to think of this degradation is to calculate on the assumption that the signal-to-noise is set by the photon shot noise but that the detector has half the detective quantum efficiency that it has been for mode 1 above.

3. The CCD is operated with high gain in the multiplication register so that the readout noise of the CCD output amplifier is completely negligible for each multiplied electron. If each event is then thresholded and treated as a single event on equal weight without making any attempt to consider its amplitude then the quantum efficiency that is lost by operating in mode 2 above is restored, giving the same quantum efficiency essentially as that in mode 1 above. There are, however, the major limitations that the maximum photon rate used must be kept extremely low in order to avoid coincidence losses which will give rise to non-linearities in the response curve of the detector system, and the corresponding need for deep cooling to maintain correspondingly low levels of dark current.

The fact that it is possible to change the mode of operation, or to design systems which work in modes 2 and 3 above simultaneously so easily offers a great deal of flexibility.

There can be little doubt that the new technology developed by Marconi Applied Technologies will have a substantial impact on the design of a wide range of scientific imaging systems. The ability demonstrated here to operate a CCD system with essentially no readout noise and yet for it not to affect the performance of the CCD detector in other ways is quite remarkable and quite unique.